\documentclass[11pt(*)]{JHEP3}
\usepackage{appendix}
\usepackage{amsmath}
\usepackage{graphicx}
\usepackage{amssymb}
\usepackage{amsthm}
\usepackage{latexsym}
\usepackage{mathrsfs}
\usepackage{epsfig,multicol,bbm}
\usepackage{slashbox}
\usepackage[square, comma, sort&compress]{natbib}

\newcommand\fverb{\setbox\fverbbox=\hbox\bgroup\verb}
\newcommand\fverbdo{\egroup\medskip\noindent%
            \fbox{\unhbox\fverbbox}\ }
\newcommand\fverbit{\egroup\item[\fbox{\unhbox\fverbbox}]}
\newbox\fverbbox

\title{Mass Matrices and Their Renormalization}
\author{Shao-Hsuan Chiu\\
Chang Gung University, Tao-Yuan
TAIWAN, Republic of China.\\
E-mail:\email{schiu@mail.cgu.edu.tw
}}

\author{
T. K. Kuo\\
Purdue University, Department of Physics, West Lafayette,
IN 47906, USA.\\
E-mail:\email{tkkuo@physics.purdue.edu}}

\author{Tae-Hun Lee\\
   Purdue University, Department of Physics, West Lafayette,
IN 47906, USA.\\
    E-mail: \email{lee109@physics.purdue.edu}}

\author{Chi Xiong\\
   University of Virginia, Department of Physics,
Charlottesville, VA 22904-4714, USA.\\
    E-mail: \email{cx4d@Virginia.edu}}

\received{\today}
\accepted{\today}
\preprint{\hepph{arXiv:0811.1806}}  

\abstract{We obtain explicitly the renormalization group equations for the quark mass matrices in terms of a set of rephasing invariant parameters. For a range of assumed high energy values for the mass ratios and mixing parameters, they are found to evolve rapidly and develop hierarchies as the energy scale decreases. To achieve the experimentally observed high degree of hierarchy, however, the introduction of new models with specific properties becomes necessary.}


\dedicated{}

\begin{document}
\section{Introduction}

A long-standing problem in particle physics is to have some theoretical insights regarding the plethora of parameters contained in the mass matrices of quarks and leptons. These matrices arise from the coupling of the Higgs bosons to the fermions, and are not constrained by any known principle. The coupling gives rise to ten parameters each in either the quark or the lepton sector. These seemingly arbitrary quantities do have some regularities. In the quark sector, both the mass ratios and mixing parameters exhibit rather large hierarchies. On the other hand, while the lepton masses are hierarchical, the neutrino masses tend to be more degenerate with some large mixing angles. When one tries to sort out possible clues contained therein, we need also keep in mind that these parameters are all measured at low energies, and it is necessary to bring renormalization effects into the picture. As the energy scale changes, one expects the pattern of regularity to evolve according to the renormalization group equations(RGE). Thus, we might entertain the hope that a simpler picture will emerge at high energies. The analysis of the RGE of the quark mass matrices has a rich literature and a long history \cite{Cheng:1973nv,Ma:1979cw,Pendleton:1980as,Hill:1980sq,Machacek:1983fi,Sasaki:1986jv,Babu:1987im,Olechowski:1990bh,Barger:1992ac}. However, general conclusions are not easily available. For one thing, the RGE are simple when formulated in full mass matrices. This means that there are a large number of superfluous degrees of freedom which must be stripped away to get at the physical variables. However, the choice of these variables are not unique. In the literature, the extant RGE turn out to be very complicated and highly nonlinear. As a result, while their low energy behavior is generally known, very little can be said of their intermediate and/or high energy behaviors.

In this paper we propose to write the RGE in terms of a set of rephasing invariant mixing parameters which were introduced recently. The resulting equations are simpler than those given in terms of other parameters and are amenable to a general analysis. At low energies, it is found that the RGE are close to a fixed point when we put in the physical values observed experimentally. This result is well-known in the literature, so there is very little evolution of the parameters in the low energy region. If one were to extrapolate this low energy behavior to all energies, then one might conclude that renormalization effects are altogether unimportant. However, this extrapolation is based on the assumption that the RGE are exact for all energies and that the initial values are precisely known. A more likely scenario is that what we have at low energies is an effective theory which comes from some new theory valid at high energies. It is therefore more appropriate to evolve the RGE from high to low energies. To do this we will assume that the parameters are generic, instead of being hierarchical, at high energies. It is seen that the set of RGE does give rapid running for a range of initial values. This suggests an ``infrared hierarchy'' scenario, viz., low  energy hierarchy comes from renormalization evolution for a range of generic values at high energies. To account for the very large hierarchy (up to $e^{-12}\sim\lambda^8$, $\lambda\simeq0.23$) observed experimentally, however, it is necessary to invoke new theories different from the standard model(SM) or its minimal supersymmetric extension(MSSM). In the absence of concrete models we mimic their effects by changing some parameter in the known RGE. It is found that, for appropriate choices, all mixing parameters evolve very rapidly and infrared hierarchy is a viable scenario. However, mass ratios do   not develop a strong enough hierarchy to match the observed values. Obviously, more work needs to be done and it is necessary to construct explicit models to realize these ideas. We hope to report on our progress in a future publication.

This paper is organized as follows. In Sec.\ref{Sec:Rephasing Invariant}, we review and summarize the $(x,y)$ parametrization introduced earlier. The RGE for $(x,y)$ and mass ratios are given in Sec. \ref{Sec:RGE}. Section-\ref{Sec:Numerical Results} is devoted to the exploration of solutions for these equations which are compatible with the idea of infrared hierarchy. After some concluding remarks in Sec. \ref{Sec:Conclusion}, we discuss in Appendix \ref{Appx:AiBi} detailed properties of the matrices which enter the RGE. Finally, brief summaries on two-loop renormalization are included in Appendix \ref{Appx:two-loop RGE}.
\section{Rephasing Invariant Parametrization}\label{Sec:Rephasing Invariant}

To establish our notation and for completeness, we now review briefly the rephasing invariant parametrization introduced earlier \cite{Kuo:2005pf, Kuo:2005jt}.

Three flavor mixing is described in terms of a $3\times3$ unitary matrix $V_{ij}$, where $i,j=1,2,3$. Without loss of generality, we may impose the condition $\det V=+1$. Then, there are six rephasing invariant combinations
\begin{equation}
\Gamma_{ijk}=V_{1i}V_{2j}V_{3k}=R_{ijk}-iJ,\label{eq:Gamma}
\end{equation}
 where $(i,j,k)=$ cyclic permutation of $(1,2,3)$ and $J$ is the Jarlskog invariant \cite{Jarlskog:1985ht}, which turns out to be the common imaginary part of all $\Gamma_{ijk}$. We define
 \begin{equation}
 (x_1,x_2,x_3;y_1,y_2,y_3)=(R_{123}, R_{231}, R_{312};R_{132}, R_{213}, R_{321})\label{eq:(x,y)=R}
 \end{equation}
 It is found that they satisfy two constraints
 \begin{equation}
 \det V=(x_1+x_2+x_3)-(y_1+y_2+y_3)=1,\label{eq:xy-constraint-I}
 \end{equation}
 \begin{equation}
 x_1x_2+x_2x_3+x_3x_1=y_1y_2+y_2y_3+y_3y_1,\label{eq:xy-constraint-II}
 \end{equation}
 leaving four independent parameters to describe the mixing. In addition, there is a simple relation
 \begin{equation}
 J^2=x_1x_2x_3-y_1y_2y_3.\label{eq:J}
 \end{equation}
 These parameters are all bounded by $\pm1$, i.e.,
 \begin{equation}
-1\leq(x_i,y_j)\leq1,~~~y_j\leq x_i.\label{eq:xy-range}
\end{equation}
 They are related to $|V_{ij}|^2$ by
 \begin{equation}
\begin{array}{ccl}
W&=&\left(\begin{array}{ccc}
|V_{11}|^2&|V_{12}|^2&|V_{13}|^2\\
|V_{21}|^2&|V_{22}|^2&|V_{23}|^2\\
|V_{31}|^2&|V_{32}|^2&|V_{33}|^2
\end{array}\right)
=\left(\begin{array}{ccc}
x_1-y_1&x_2-y_2&x_3-y_3\\
x_3-y_2&x_1-y_3&x_2-y_1\\
x_2-y_3&x_3-y_1&x_1-y_2
\end{array}\right).
\end{array}\label{eq:W}
\end{equation}
Also, the matrix of the cofactors of $W$, with $w^TW=(\det W)I$, is given by
\begin{equation}
w=
\left(\begin{array}{ccc}
x_1+y_1&x_2+y_2&x_3+y_3\\
x_3+y_2&x_1+y_3&x_2+y_1\\
x_2+y_3&x_3+y_1&x_1+y_2
\end{array}\right).\label{eq:w}
\end{equation}
The matrix $w$ appears repeatedly in the RGE which will be presented in the following section.
Experimentally, the elements of the quark mixing matrix $V_{CKM}$ are well measured and they exhibit striking hierarchies:
\begin{equation}
(x_1,x_2,x_3;-y_1,-y_2,|y_3|)\cong(1,\lambda^6,\lambda^6;+\lambda^4,+\lambda^2,\lambda^8).\label{eq:xy-lambda}
\end{equation}
It should be emphasized that, unlike the usual CKM parametrizations, which contain terms of order $\lambda$, here the hierarchy is in powers of $\lambda^2$ only, which is also the hierarchy pattern in quark mass ratios. We add that, since we can choose any set of four of the $(x,y)$ variables to parametrize $V_{CKM}$, including $y_3$ (which is very small) in such a set implies that the physical $V_{CKM}$ is effectively described by three parameters only. Thus, there are correlations amongst the familiar four parameter sets in $V_{CKM}$. These approximate relations \cite{Kuo:2005jt} are
\begin{equation}
\begin{array}{rcl}
\rho&\cong&\rho^2+\eta^2,\\
s_{13}&\cong& c_{\delta}s_{12}s_{23},\\
|V_{us}|^2|V_{cb}|^2&\cong&|V_{td}|^2+|V_{ub}|^2.
\end{array}\label{eq:appx relation-CKM}
\end{equation}
All of these are well satisfied by existing data \cite{Amsler:2008zz}.
\section{Renormalization Group Equations}\label{Sec:RGE}

The RGE for quark mass matrices have been obtained and studied for a long time. Although they are simple when expressed in matrix form, they become complicated and highly nonlinear if one writes them in terms of the parameters commonly used for the CKM matrix. In fact, their complexity has been the major obstacle preventing one from drawing general conclusions so that most analyses are confined to low energies and their extrapolations \cite{Sasaki:1986jv, Babu:1987im, Olechowski:1990bh, Barger:1992ac}.

In this paper we will reformulate the problem using the $(x,y)$ parameters summarized in the previous section. The resulting equations turn out to be simpler and not so formidable. In addition, when we write down the evolution equations for some simple functions of the mass ratios, they take very similar forms to those of the $(x,y)$ parameters. This bolsters the idea that the observed hierarchies in the mass ratios and the $(x,y)$ parameters, which so closely resemble each other, may both be related to renormalization.

We begin by citing the one loop RGE for the mass (squared) matrix of the u-type quarks, $M_u= Y_uY_u^\dagger$, and that of the d-type quarks, $M_d=Y_dY_d^\dagger$, where $Y$ denotes the Yukawa coupling matrices of the Higgs boson to the quarks \cite{Machacek:1983fi, Sasaki:1986jv, Babu:1987im}.
\begin{equation}
\mathscr{D}M_u=a_uM_u+bM_u^2+c\{M_u,M_d\},\label{eq:DMu}
\end{equation}
\begin{equation}
\mathscr{D}M_d=a_dM_d+bM_d^2+c\{M_u,M_d\}.\label{eq:DMd}
\end{equation}
Here, $\mathscr{D}=16\pi^2\frac{d}{dt}$ and $t=\ln(\mu/M_W)$, where $\mu$ is an energy scale and $M_W$ is the $W$ boson mass. The values $(a_u,a_d,b,c)$ are model-dependent and we list them in Table \ref{t:coefficient-RGE}, for the SM and the MSSM with $\tan\beta=1$, as well as the two Higgs model(THM), where one Higgs couples to u-type quarks, and the other to d-type quarks and the leptons. Here, the notations used in Table \ref{t:coefficient-RGE} are as follow,
\begin{equation}
\begin{array}{ccl}
G_u&=&\frac{17}{20}g^2_1+\frac{9}{4}g^2_2+8g^2_3,~~ G_d=\frac{1}{4}g^2_1+\frac{9}{4}g^2_2+8g^2_3,\\
G^s_u&=&\frac{13}{15}g^2_1+3g^2_2+\frac{16}{3}g^2_3,~ G^s_d=\frac{7}{15}g^2_1+3g^2_2+\frac{16}{3}g^2_3,\\
T&=&\mbox{Tr}(3M_u+3M_d+M_e),
\end{array}
\end{equation}
where $g_i$, $i=1,2,3$, are the usual gauge coupling constants.
\TABLE[h!]{
\begin{tabular}{|c|c|c|c|c|}
	\hline
 &  $c$    &   $b$  & $a_u$   & $a_d$\\
	\hline
SM   & $-3/2$& 3 &$2(T-G_u)$ &$2(T-G_d)$\\
\hline
MSSM & +1 & 6 & $2(3\textrm{Tr}M_u-G^s_u)$ & $2(3\textrm{Tr}M_d+\textrm{Tr}M_e-G^s_d)$\\
\hline
THM & $+1/2$ & 3 & $2(3\textrm{Tr}M_u-G_u)$  &$2(3\textrm{Tr}M_d+\textrm{Tr}M_e-G_d)$\\
	\hline
\end{tabular}
\caption{Coefficients in Eqs.(\ref{eq:DMu}, \ref{eq:DMd}).}
\label{t:coefficient-RGE}}

For our purposes it is convenient to start with the explicit evolution equations of the eigenvalues of the mass matrices and those of the CKM matrix elements \cite{Sasaki:1986jv, Babu:1987im}, using Eqs.(\ref{eq:DMu}, \ref{eq:DMd}). For the eigenvalues, we have
\begin{equation}
\mathscr{D}f^2_i=f^2_i[a_u+bf^2_i+2c\sum_jh^2_j|V_{ij}|^2],\label{eq:Dg}
\end{equation}
and
\begin{equation}
\mathscr{D}h^2_j=h^2_j[a_d+bh^2_j+2c\sum_if^2_i|V_{ij}|^2].\label{eq:Df}
\end{equation}
where $f^2_i$ and $h^2_j$ denote the eigenvalues of $M_u$ and $M_d$, respectively. The explicit equations for the CKM matrix elements are
\begin{equation}
\mathscr{D}V_{ij}=c\left[\sum_{\ell,k\neq i}F_{ik}h^2_\ell V_{i\ell}V^*_{k\ell}V_{kj}
+\sum_{m,k\neq j}H_{jk}f^2_mV^*_{mk}V_{mj}V_{ik}\right],\label{eq:DVij}
\end{equation}
where we have defined
\begin{equation}
F_{ik}=\frac{f^2_i+f^2_k}{f^2_i-f^2_k},~~~
H_{jk}=\frac{h^2_j+h^2_k}{h^2_j-h^2_k}.\label{eq:GF}
\end{equation}

Eq.(\ref{eq:DVij}), as it stands, is not rephasing invariant. To it one could append terms with $i=k$ (or $j=k$) on the right hand side \cite{Sasaki:1986jv, Kielanowski:2008wm}. They come from (purely imaginary) diagonal elements of ($UdU^\dagger/dt$), where $U$ diagonalizes $M_u$(or $M_d$). A rephasing transformation, $U\rightarrow(\exp i\alpha(t))U$, where $\alpha(t)$ is a $t$-dependent, diagonal, phase matrix, yields $(UdU^\dagger/dt)\rightarrow e^{i\alpha}(UdU^\dagger/dt)e^{-i\alpha}-id\alpha/dt$. Thus, the diagonal elements of ($UdU^\dagger/dt$) are rephasing dependent, and Eq.(\ref{eq:DVij}) should only be used to compute the evolution of rephasing invariant combinations of $V_{ij}$, such as $|V_{ij}|^2$ or $(x_i,y_j)$.

Using these equations, while keeping in mind that the condition $\det V=+1$ implies relations such as $V^*_{11}=V_{22}V_{33}-V_{23}V_{32}$, etc., we can work out the RGE for the mass ratios and the mixing parameters in terms of the $(x,y)$ parameters. We find, after some algebra,
\begin{equation}
-\mathscr{D}x_i/c=(\Delta f_{23}, \Delta f_{31}, \Delta f_{12})A_i\left(\begin{array}{c}H_{23}\\H_{31}\\H_{12}\end{array}\right)
+(\Delta h_{23}, \Delta h_{31}, \Delta h_{12})B_i\left(\begin{array}{c}F_{23}\\F_{31}\\F_{12}\end{array}\right),
\label{eq:Dx}
\end{equation}
\begin{equation}
-\mathscr{D}y_i/c=(\Delta f_{23}, \Delta f_{31}, \Delta f_{12})A^\prime_i\left(\begin{array}{c}H_{23}\\H_{31}\\H_{12}\end{array}\right)
+(\Delta h_{23}, \Delta h_{31}, \Delta h_{12})B^\prime_i\left(\begin{array}{c}F_{23}\\F_{31}\\F_{12}\end{array}\right),
\label{eq:Dy}
\end{equation}
where  we have defined the differences of mass squared,
\begin{equation}
\Delta f_{ij}=f^2_i-f^2_j,~~~\Delta h_{ij}=h^2_i-h^2_j.\label{eq:def-mass-diffs}
\end{equation}
The matrices $A_i$, $B_i$, $A^\prime_i$ and $B^\prime_i$ are given in Table \ref{t:AiBi}. Note that all of the elements of these matrices are bounded [$0\leq(|x_ix_j|,|x_iy_j|,|y_iy_j|)\leq1$]. As will be discussed in detail in Appendix \ref{Appx:AiBi}, these matrices satisfy consistency relations which guarantee the identities
\begin{equation}
\mathscr{D}\sum (x_i-y_i)=0,~~\mathscr{D}\sum_{i<j}(x_ix_j-y_iy_j)=0.
\end{equation}
Also, by calculating $\mathscr{D}(x_1x_2x_3)$ and $\mathscr{D}(y_1y_2y_3)$ (see Eqs.(\ref{eq:identity-xyAJ}, \ref{eq:identity-xyBJ})), we obtain a simple evolution equation for $J^2$:
\begin{equation}
\mathscr{D}J^2=-2cJ^2(\Delta f^T\cdot w\cdot H+\Delta h^T\cdot w^T\cdot F),\label{eq:DJ}
\end{equation}
where we have used a compact notation for the column matrices:
\begin{equation}
\{\Delta f,\Delta h, F, H\}
=\left\{
\left(\begin{array}{c}
\Delta f_{23}\\
\Delta f_{31}\\
\Delta f_{12}
\end{array}\right),
\left(\begin{array}{c}
\Delta h_{23}\\
\Delta h_{31}\\
\Delta h_{12}
\end{array}\right),
\left(\begin{array}{c}
F_{23}\\
F_{31}\\
F_{12}
\end{array}\right),
\left(\begin{array}{c}
H_{23}\\
H_{31}\\
H_{12}
\end{array}\right)
\right\}.\label{eq:notations-fhFH}
\end{equation}
Eq.(\ref{eq:DJ}) agrees with previous results.

\newpage
\TABLE[h]{
\begin{tabular}{|c||c|c|}
	\hline
\backslashbox{$i$}{}&$A_i$&$B_i$\\
\hline\hline
&&\\
1&$x_1\left(\begin{array}{ccl}
              y_1& x_2&x_3\\
              x_3&y_3&x_2\\
              x_2&x_3&y_2
            \end{array}\right)
            +\left(\begin{array}{ccl}
            y_1x_1&y_3y_2&y_2y_3\\
            y_1y_2&y_3x_1&y_2y_1\\
            y_1y_3&y_3y_1&y_2x_1
\end{array}\right)$&$x_1\left(\begin{array}{ccl}
              y_1& x_3&x_2\\
              x_2&y_3&x_3\\
              x_3&x_2&y_2
            \end{array}\right)
            +\left(\begin{array}{ccl}
            y_1x_1&y_3y_2&y_2y_3\\
            y_1y_2&y_3x_1&y_2y_1\\
            y_1y_3&y_3y_1&y_2x_1
\end{array}\right)$\\
&&\\
\hline
&&\\
2&$x_2\left(\begin{array}{ccl}
              x_1& y_2&x_3\\
              x_3&x_1&y_1\\
              y_3&x_3&x_1
            \end{array}\right)
            +\left(\begin{array}{ccl}
            y_3y_1&y_2x_2&y_1y_3\\
            y_3y_2&y_2y_3&y_1x_2\\
            y_3x_2&y_2y_1&y_1y_2
\end{array}\right)$&$x_2\left(\begin{array}{ccl}
              x_1& x_3&y_3\\
              y_2&x_1&x_3\\
              x_3&y_1&x_1
            \end{array}\right)
            +\left(\begin{array}{ccl}
            y_2y_1&y_1y_2&y_3x_2\\
            y_2x_2&y_1y_3&y_3y_1\\
            y_2y_3&y_1x_2&y_3y_2
\end{array}\right)$\\
&&\\
\hline
&&\\
3&$x_3\left(\begin{array}{ccl}
              x_1& x_2&y_3\\
              y_2&x_1&x_2\\
              x_2&y_1&x_1
            \end{array}\right)
            +\left(\begin{array}{ccl}
            y_2y_1&y_1y_2&y_3x_3\\
            y_2x_3&y_1y_3&y_3y_1\\
            y_2y_3&y_1x_3&y_3y_2
\end{array}\right)$&$x_3\left(\begin{array}{ccl}
              x_1& y_2&x_2\\
              x_2&x_1&y_1\\
              y_3&x_2&x_1
            \end{array}\right)
            +\left(\begin{array}{ccl}
            y_3y_1&y_2x_3&y_1y_3\\
            y_3y_2&y_2y_3&y_1x_3\\
            y_3x_3&y_2y_1&y_1y_2
\end{array}\right)$\\
&&\\
\hline\hline
\backslashbox{$i$}{}&$A^\prime_i$&$B^\prime_i$\\
\hline\hline
&&\\
1&$y_1\left(\begin{array}{ccl}
              x_1& y_2&y_3\\
              y_2&y_3&x_2\\
              y_3&x_3&y_2
            \end{array}\right)
            +\left(\begin{array}{ccl}
            x_1y_1&x_3x_2&x_2x_3\\
            x_1x_3&x_3x_1&x_2y_1\\
            x_1x_2&x_3y_1&x_2x_1
\end{array}\right)$&
$y_1\left(\begin{array}{ccl}
              x_1& y_2&y_3\\
              y_2&y_3&x_3\\
              y_3&x_2&y_2
            \end{array}\right)
            +\left(\begin{array}{ccl}
            x_1y_1&x_2x_3&x_3x_2\\
            x_1x_2&x_2x_1&x_3y_1\\
            x_1x_3&x_2y_1&x_3x_1
\end{array}\right)$\\
&&\\
\hline
&&\\
2&$y_2\left(\begin{array}{ccl}
              y_1& x_2&y_3\\
              x_3&y_3&y_1\\
              y_3&y_1&x_1
            \end{array}\right)
            +\left(\begin{array}{ccl}
            x_3x_1&x_2y_2&x_1x_3\\
            x_3y_2&x_2x_1&x_1x_2\\
            x_3x_2&x_2x_3&x_1y_2
\end{array}\right)$&
$y_2\left(\begin{array}{ccl}
              y_1& x_3&y_3\\
              x_2&y_3&y_1\\
              y_3&y_1&x_1
            \end{array}\right)
            +\left(\begin{array}{ccl}
            x_2x_1&x_3y_2&x_1x_2\\
            x_2y_2&x_3x_1&x_1x_3\\
            x_2x_3&x_3x_2&x_1y_2
\end{array}\right)$\\
&&\\
\hline
&&\\
3&$y_3\left(\begin{array}{ccl}
              y_1& y_2&x_3\\
              y_2&x_1&y_1\\
              x_2&y_1&y_2
            \end{array}\right)
            +\left(\begin{array}{ccl}
            x_2x_1&x_1x_2&x_3y_3\\
            x_2x_3&x_1y_3&x_3x_2\\
            x_2y_3&x_1x_3&x_3x_1
\end{array}\right)$&
$y_3\left(\begin{array}{ccl}
              y_1& y_2&x_2\\
              y_2&x_1&y_1\\
              x_3&y_1&y_2
            \end{array}\right)
            +\left(\begin{array}{ccl}
            x_3x_1&x_1x_3&x_2y_3\\
            x_3x_2&x_1y_3&x_2x_3\\
            x_3y_3&x_1x_2&x_2x_1
\end{array}\right)$\\
&&\\
\hline
\end{tabular}
\caption{The matrices $A_i$, $B_i$, $A^\prime_i$ and $B^\prime_i$ used in Eqs.(\ref{eq:Dx}, \ref{eq:Dy}).}\label{t:AiBi}
}
\newpage
We now turn to the RGE of mass ratios. From Eqs.(\ref{eq:Dg}, \ref{eq:Df}), it is straightforward to obtain
\begin{equation}
\mathscr{D}\left(\begin{array}{c}
\ln R_{23}\\
\ln R_{31}\\
\ln R_{12}
\end{array}\right)=
\frac{b}{2}\left(\begin{array}{c}
\Delta f_{23}\\
\Delta f_{31}\\
\Delta f_{12}
\end{array}\right)
+c\left(\begin{array}{ccl}
x_1+y_1&x_2+y_2&x_3+y_3\\
x_3+y_2&x_1+y_3&x_2+y_1\\
x_2+y_3&x_3+y_1&x_1+y_2
\end{array}\right)
\left(\begin{array}{c}
\Delta h_{23}\\
\Delta h_{31}\\
\Delta h_{12}
\end{array}\right),\label{eq:DlnR}
\end{equation}
\begin{equation}
\mathscr{D}\left(\begin{array}{c}
\ln r_{23}\\
\ln r_{31}\\
\ln r_{12}
\end{array}\right)=
\frac{b}{2}\left(\begin{array}{c}
\Delta h_{23}\\
\Delta h_{31}\\
\Delta h_{12}
\end{array}\right)
+c\left(\begin{array}{ccl}
x_1+y_1&x_3+y_2&x_2+y_3\\
x_2+y_2&x_1+y_3&x_3+y_1\\
x_3+y_3&x_2+y_1&x_1+y_2
\end{array}\right)
\left(\begin{array}{c}
\Delta f_{23}\\
\Delta f_{31}\\
\Delta f_{12}
\end{array}\right).\label{eq:Dlnr}
\end{equation}
Here, we have defined ratios of masses as
\begin{equation}
R_{ij}=f_i/f_j,~~~r_{ij}=h_i/h_j.\label{eq:def-mass-ratios}
\end{equation}
We summarize this by writing Eqs.(\ref{eq:DlnR},\ref{eq:Dlnr}) in the form
\begin{equation}
\mathscr{D}\ln R=\frac{b}{2}\Delta f+cw\Delta h,
\end{equation}
\begin{equation}
\mathscr{D}\ln r=\frac{b}{2}\Delta h+cw^T\Delta f,
\end{equation}
where we used Eq.(\ref{eq:notations-fhFH}) and the definition
\begin{equation}
(\ln R,\ln r)
=\left\{
\left(\begin{array}{c}
\ln R_{23}\\
\ln R_{31}\\
\ln R_{12}
\end{array}\right),
\left(\begin{array}{c}
\ln r_{23}\\
\ln r_{31}\\
\ln r_{12}
\end{array}\right)
\right\}.
\end{equation}

In going from eigenvalues ($f_i,h_i$) to their ratios, we find a set of much simplified RGE which only depends on the mass differences and the ($x,y$) parameters in the combination $w$ and $w^T$, defined in Eq.(\ref{eq:w}), a fact which seems very interesting but not understood. The RGE for the mass ratios, Eqs.(\ref{eq:DlnR}, \ref{eq:Dlnr}), can be brought into a form in conformity with those for the $(x,y)$ parameters by considering $\sinh(\ln R_{ij})=(f^2_i-f^2_j)/2f_if_j$, with $F_{ij}=\coth(\ln R_{ij})$. Note that in the hierarchical limit $\sinh(\ln R_{ij})\rightarrow \frac{1}{2}(f_i/f_j)$, for $f_i\gg f_j$. Let us define
\begin{equation}
Q_{ij}=\ln[\sinh(\ln R_{ij})],~~~q_{ij}=\ln[\sinh(\ln r_{ij})].\label{eq:def-Q-q}
\end{equation}
We find
\begin{equation}
\begin{array}{ccl}
\mathscr{D}(Q_{23},Q_{31},Q_{12})&=&\frac{b}{2}(\Delta f_{23},\Delta f_{31},\Delta f_{12})\left(\begin{array}{ccl}
F_{23}&0&0\\
0&F_{31}&0\\
0&0&F_{12}
\end{array}\right)\\\\
&&+c(\Delta h_{23},\Delta h_{31},\Delta h_{12})w^T\left(\begin{array}{ccl}
F_{23}&0&0\\
0&F_{31}&0\\
0&0&F_{12}
\end{array}\right),
\end{array}\label{eq:DQ}
\end{equation}
\\
\begin{equation}
\begin{array}{ccl}
\mathscr{D}(q_{23},q_{31},q_{12})&=&\frac{b}{2}(\Delta h_{23},\Delta h_{31},\Delta h_{12})\left(\begin{array}{ccl}
H_{23}&0&0\\
0&H_{31}&0\\
0&0&H_{12}
\end{array}\right)\\\\
&&+c(\Delta f_{23},\Delta f_{31},\Delta f_{12})w\left(\begin{array}{ccl}
H_{23}&0&0\\
0&H_{31}&0\\
0&0&H_{12}
\end{array}\right).
\end{array}\label{eq:Dq}
\end{equation}
The evolutions of $Q_{ij}$ and $q_{ij}$ are thus similar in form to those of $(x_i,y_j)$ and $J^2$. In fact, using these equations we readily recover the elegant result for the evolution of the $CP-$violation measure \cite{Athanasiu:1986mk}, which is a product of $J$ and the quark mass-squared differences.
\begin{equation}
\mathscr{D}\ln [J\prod(\Delta f_{ij})\prod(\Delta h_{k\ell})/(\prod f^2_i\prod h^2_j)]=b\mbox{Tr}(M_u+M_d)
\label{eq:Identity-JgfTr}.
\end{equation}
where $b$ is given in Table \ref{t:coefficient-RGE}. Eq.(\ref{eq:Identity-JgfTr}) agrees with Ref. \cite{Athanasiu:1986mk} when we use the RGE of $\det M_u$ and $\det M_d$, given by
\begin{equation}
\mathscr{D}(\ln\det M_u)=3a_u+b\mbox{Tr}M_u+2c\mbox{Tr}M_d,\label{eq:DlndetMu}
\end{equation}
\begin{equation}
\mathscr{D}(\ln\det M_d)=3a_d+b\mbox{Tr}M_d+2c\mbox{Tr}M_u.\label{eq:DlndetMd}
\end{equation}
The ten independent equations, which are contained in Eqs.(\ref{eq:Dx}, \ref{eq:Dy}, \ref{eq:DQ}, \ref{eq:Dq}, \ref{eq:DlndetMu}, \ref{eq:DlndetMd}), form a complete set of RGE for the ten physical parameters in $M_u$ and $M_d$. Although it does not seem feasible to analytically\cite{Balzereit:1998id} disentangle these coupled, non-linear, differential equations, their dependence on the relevant variables are rather simple. So their analyses are certainly less demanding compared to the corresponding equations in the literature, which were written in terms of other parametrizations.

We offer some general remarks. Let us first observe that the RGE for the $(x,y)$ variables, Eqs.(\ref{eq:Dx}, \ref{eq:Dy}), have a fixed point at the parameter set (or permutation thereof): $x_1=1$, $x_i=y_j=0$, $i\neq1$. The physical values for $(x_i,y_j)$ are thus very close to this fixed point, and little evolution is expected of these variables within the low energy region. As for the evolution of mass ratios, note first that for hierarchical ratios, Eqs.(\ref{eq:DlnR}, \ref{eq:Dlnr}) coincide with Eqs.(\ref{eq:DQ}, \ref{eq:Dq}), with their fixed point at $\Delta f_{ij}=\Delta h_{ij}=0.$ Thus, in the physical low energy region, with $w\cong w^T\cong I$, and the only appreciable physical value being $f_3\simeq1$, again the mass ratios do not change much. (Numerically, the evolution over a region near $t=0$, say $\Delta t\simeq10$, $\mathscr{D}R_{ij}\cdot\Delta t/R_{ij}$ would be no more than $10\%$.) We can safely conclude that, in the low energy region, renormalization effects are small and little deviations are expected of the measured $(x,y)$ values and mass ratios. This conclusion, that the low energy physical values are close to a fixed point of the RGE, confirms the known numerical analyses given in the literature. One could extrapolate these results to high energies. But to do so accurately assumes that not only the RGE are exact, but also the parameters are very precisely known. We would rather examine these equations from the high energy end and study their behavior which might impact low energy physics. The structure of the RGE suggests that one can have rapid evolution if 1) masses are large; 2) the $(x,y)$ values are large; 3) degeneracy so that $F_{ij}$ and/or $H_{ij}$ are large. Under theses initial conditions it is possible to realize the ``infrared hierarchy'' scenario, namely, the low energy hierarchies originated from RGE which quickly drive the relevant quantities toward the observed values. As the energy scale decreases, the high energy theory is taken over by the low energy models. However, they will have little effect on the established hierarchies.
\section{Numerical Results}\label{Sec:Numerical Results}

In the absence of analytical solutions for the RGE, we turn to a numerical analysis of these equations. As we argued in the previous section, while the low energy physical values are close to those of a fixed point of the RGE, it is not clear how fast they are approached from presumed, more generic, high energy values. The structure of RGE suggests that rapid evolution is possible if all of the $(x_i,y_j)$ parameters are of order one and that the masses are large and nearly degenerate. In addition, the constants $b$ and $c$, which are characteristic of the underlying model (Table \ref{t:coefficient-RGE}), are important factors in determining the evolution speed.

\begin{figure}[h!]
  \begin{center}
\begin{tabular}{ccc}
\includegraphics[scale=0.5]{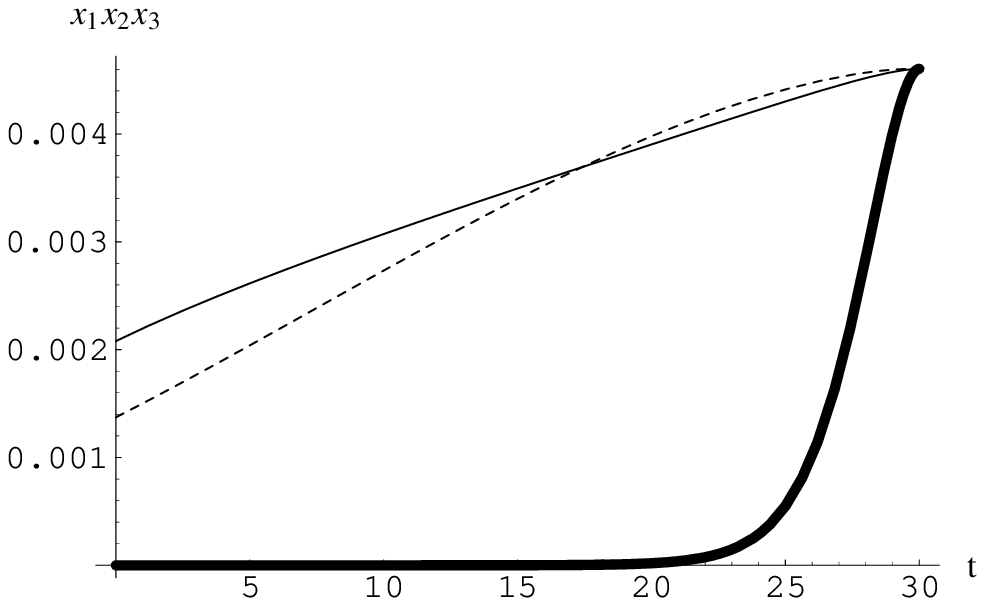} &
    \includegraphics[scale=0.5]{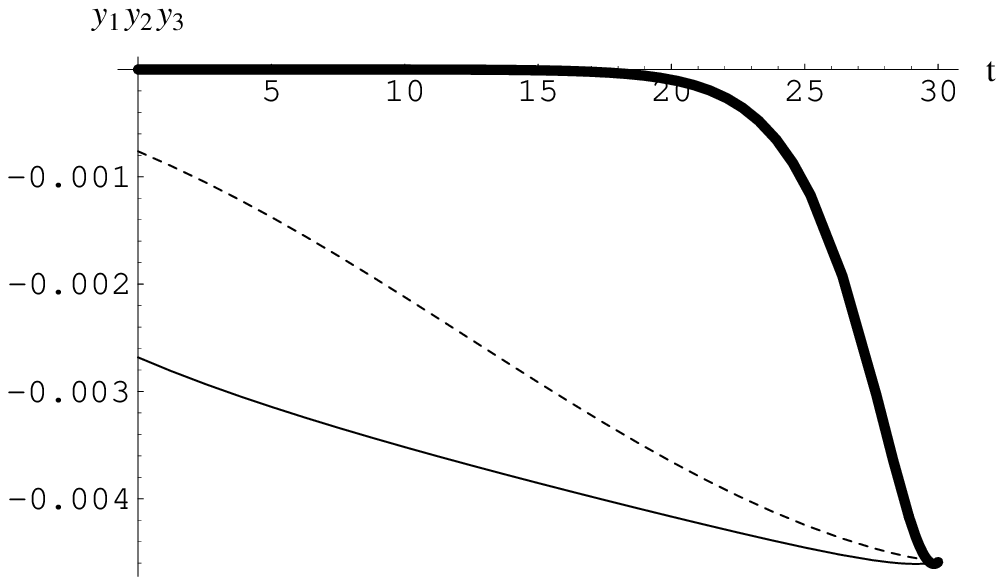}&
    \includegraphics[scale=0.5]{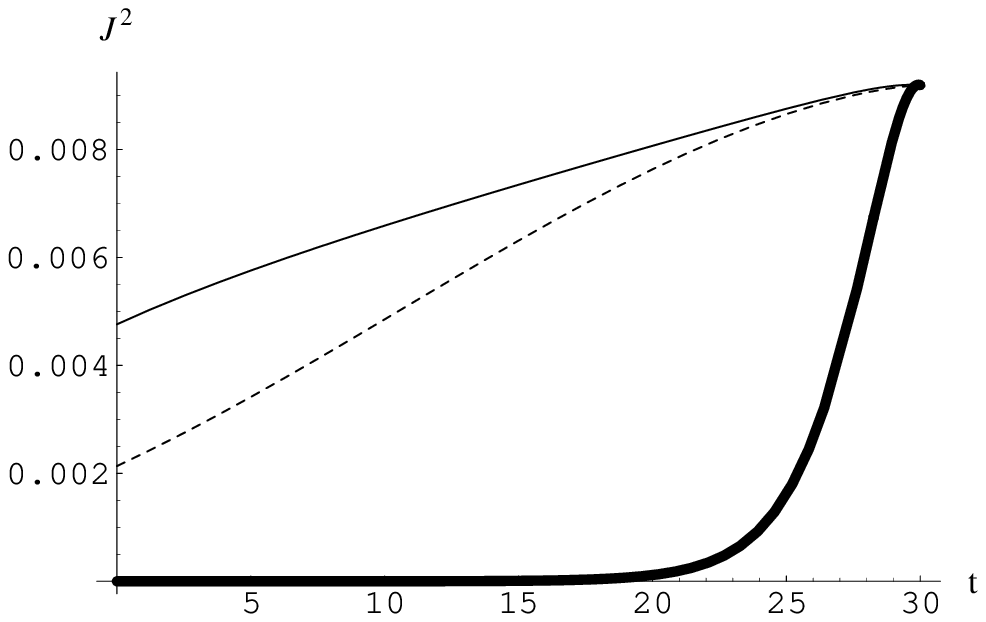}\\
    (a)&(b)&(c)
    \end{tabular}
    \end{center}
\caption{The evolutions of (a) $x_1x_2x_3$, (b) $y_1y_2y_3$ and (c) $J^2$ for SM(thin), MSSM(dashed) and the model(thick) with $(b,c)=(3,-7)$, with the boundary conditions [$(x_1,x_2)=(1/6+\epsilon,1/6-\epsilon)$,  $(y_1,y_2)=(-1/6+\epsilon,-1/6+\epsilon)$, $\epsilon=0.01$] and [$f_i=(1.6, 1.8, 2.0)$, $h_i=(0.5,0.7,0.9)$] at $t$=30.}\label{fig:xxxyyy}
\end{figure}

Before we embark on the detailed evolution of individual parameters, we will first consider the running of the quantities $x_1x_2x_3$ and $y_1y_2y_3$. They may be regarded as the analogs of phase space volumes and are measures of the overall magnitudes of the mixing parameters. We present the results in Fig.\ref{fig:xxxyyy}, where we assume that, at high energy ($t=30$, or $E\cong10^{15}\mbox{GeV})$, the initial values are close to the ``symmetric point'' $(x_i=-y_j=1/6)$. We also assume that the Higgs Yukawa coupling constants are large $(\sim\mbox{O}(1))$, and not far from being degenerate.(The graphs are not sensitive to the precise initial values chosen in presenting the plots.) These choices, apart from esthetic reasons, are also motivated by the known neutrino parameters, which are usually considered to be a reflection of high energy behavior. It is seen that, with these initial conditions, $\prod x_i$ and $\prod y_j$ drop considerably as $t$ decreases. However, the low energy physical values ($\prod x_i\simeq10^{-8}$, $\prod y_j\simeq10^{-9}$) require a drop $\sim10^{-6}$ from the symmetric point, and neither the SM or MSSM model can yield that. On the other hand, if we have some new model at high energies, it is possible that such a precipitous drop may be achieved. For this purpose we will simulate their effects by changing the values of $b$ and $c$ in Eqs.(\ref{eq:Dx}, \ref{eq:Dy}, \ref{eq:DQ}, \ref{eq:Dq}). This seems reasonable since, for the known examples, the basic structure of the RGE is the same, with the coefficients varying from one model to another. In addition, as we will see in Appendix \ref{Appx:two-loop RGE}, the two-loop RGE are also very similar and can be approximated, at least for a range of $t$ values, by using effective $b$ and $c$ values in these equations. Another example can be found in extra dimension models. Although such models are non-renormalizable, it has been argued that the effect of introducing a cutoff amounts to scale dependent couplings which can be described by equations that are just like renormalization group equations \cite{Dienes:1998vg, Bhattacharyya:2006ym}. For the Yukawa matrices, in fact, the resulting equations are similar to Eqs.(\ref{eq:DMu}, \ref{eq:DMd}), where the coefficients $b$ and $c$ become proportional to the number of Kaluza-Klein states that contribute at a given energy scale. Thus, in our attempt to assess the behavior of the RGE, we will treat $b$ and $c$ as free parameters. As we will see, it turns out that, with our choice of initial values, the evolution does not vary much as $b$ changed, but it is very sensitive to the value of $c$. Thus, it is found that, for $c\sim-7$, $\prod x_i$ and $\prod y_j$ can indeed change dramatically in the neighborhood of $t\sim30$.

\begin{figure}[h!]
  \begin{center}
\begin{tabular}{cc}
    \includegraphics[scale=0.65]{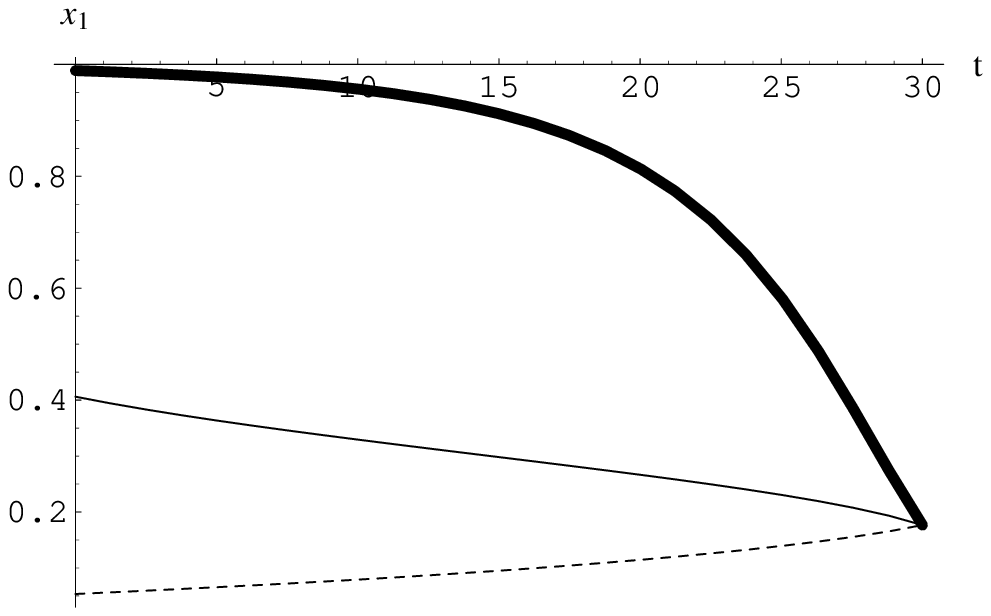} &
    \includegraphics[scale=0.65]{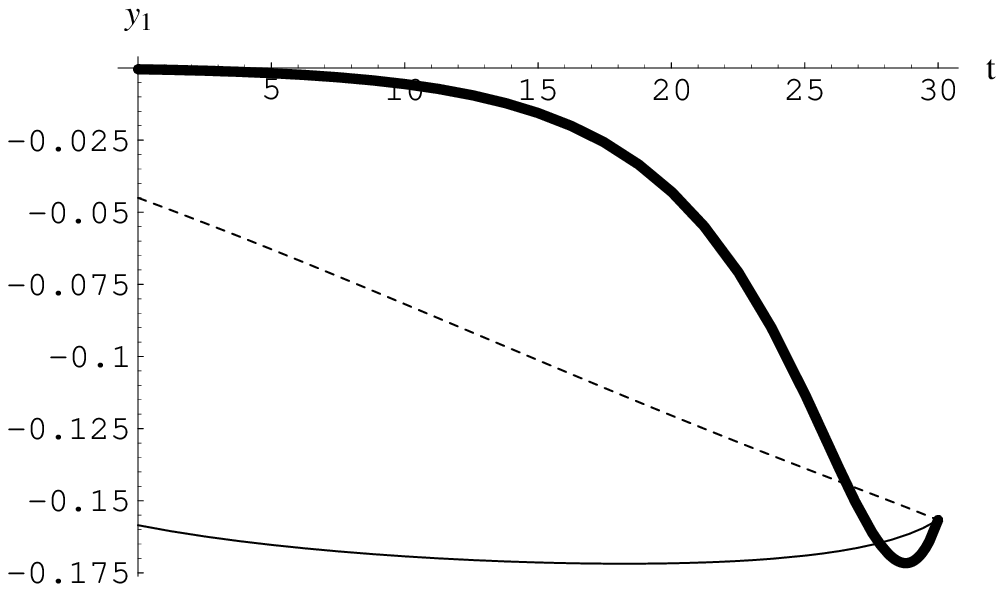} \\
    \includegraphics[scale=0.65]{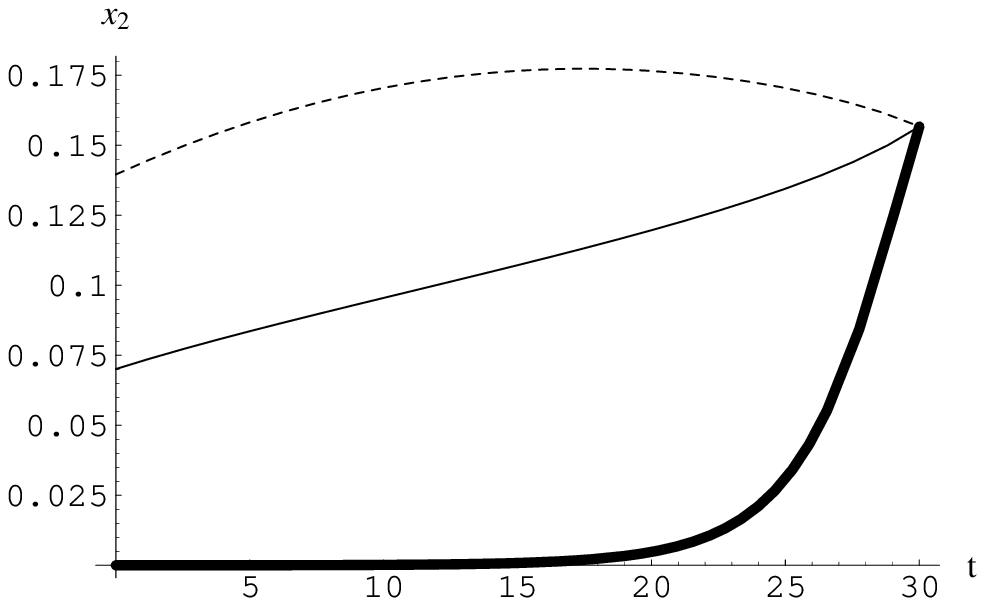} &
    \includegraphics[scale=0.65]{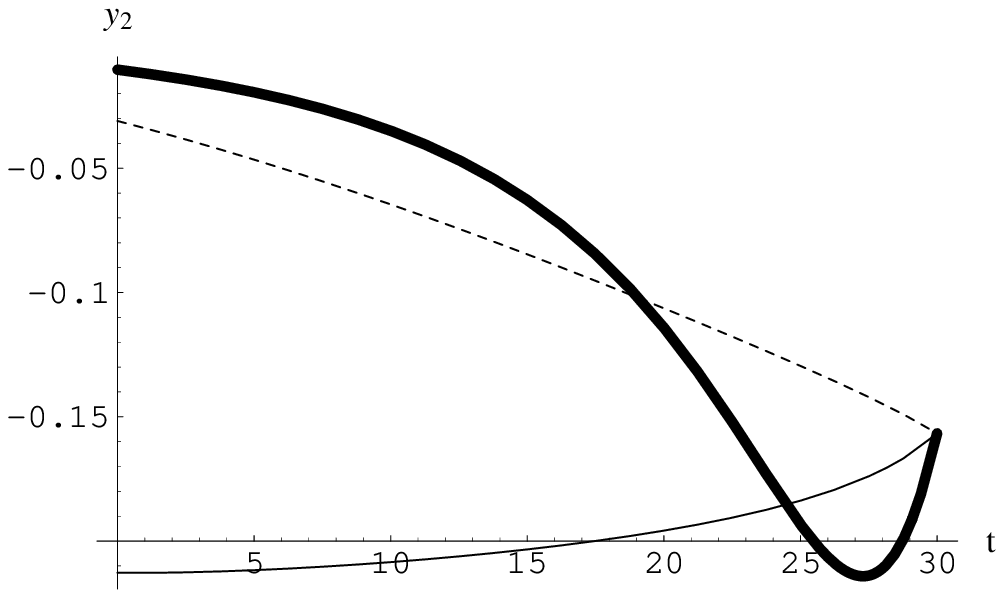} \\
    \includegraphics[scale=0.65]{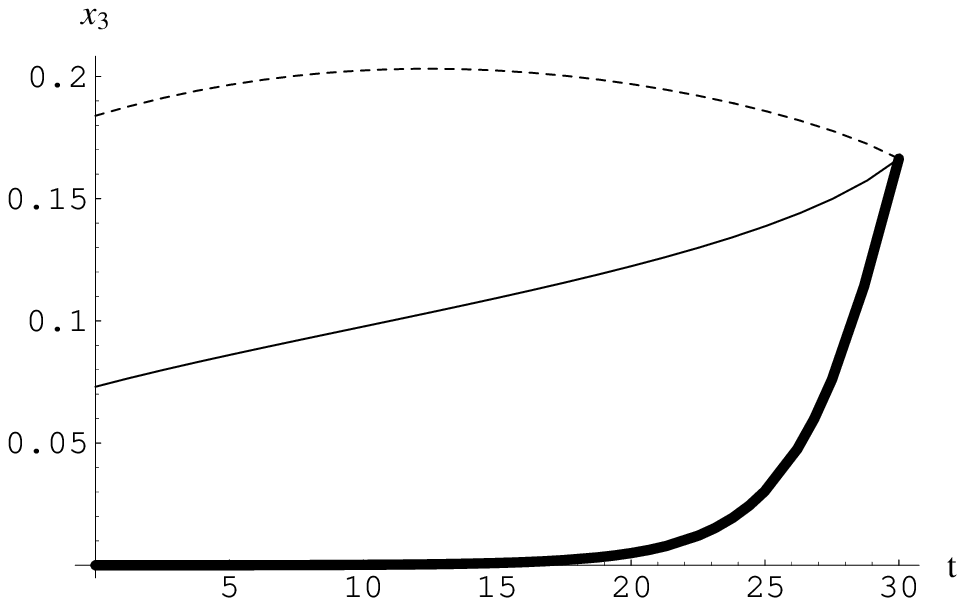} &
    \includegraphics[scale=0.65]{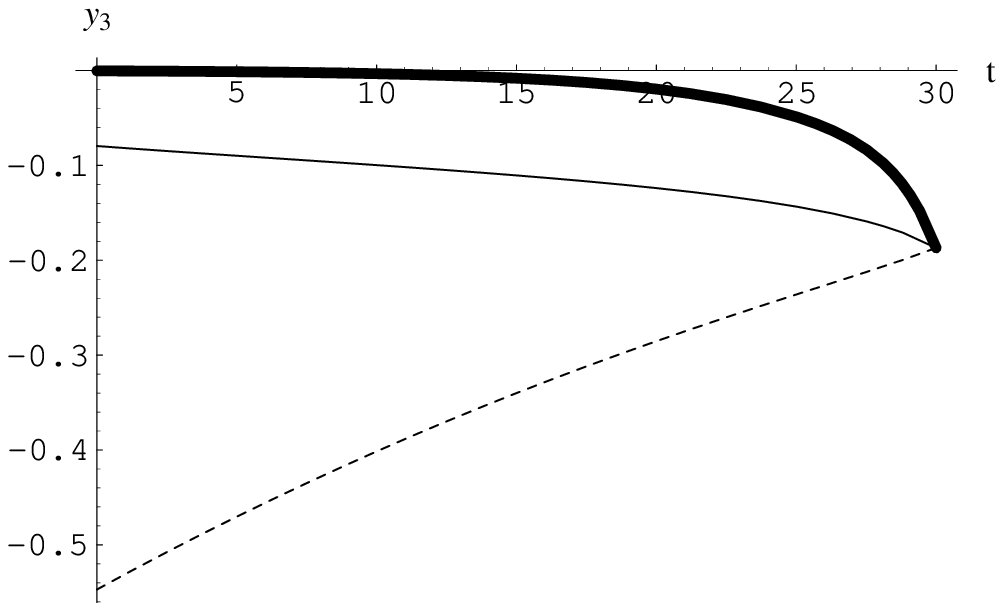} \\
    (a) & (b)
\end{tabular}
    \end{center}
\caption{The evolutions of (a) $x_i$ and (b) $y_i$ for SM(thin), MSSM(dashed) and the model(thick) with $(b,c)=(3,-7)$, with the boundary conditions [$(x_1,x_2)=(1/6+\epsilon,1/6-\epsilon)$,  $(y_1,y_2)=(-1/6+\epsilon,-1/6+\epsilon)$, $\epsilon=0.01$] and [$f_i=(1.6, 1.8, 2.0)$, $h_i=(0.5,0.7,0.9)$] at $t$=30.}\label{fig:xy}
\end{figure}

\begin{figure}[h!]
  \begin{center}
\begin{tabular}{ccc}
    \includegraphics[scale=0.7]{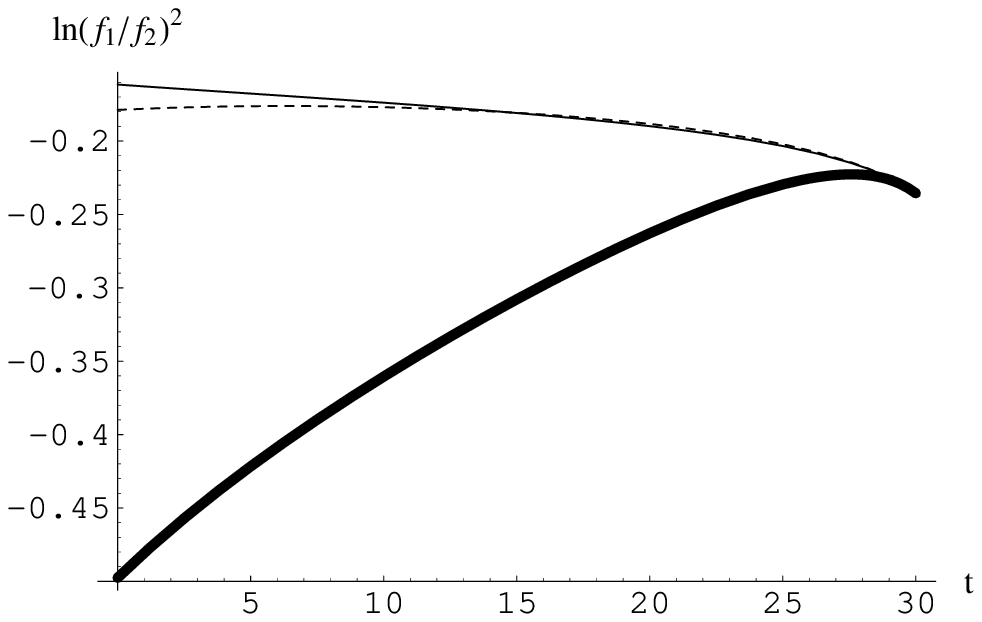} &&
    \includegraphics[scale=0.7]{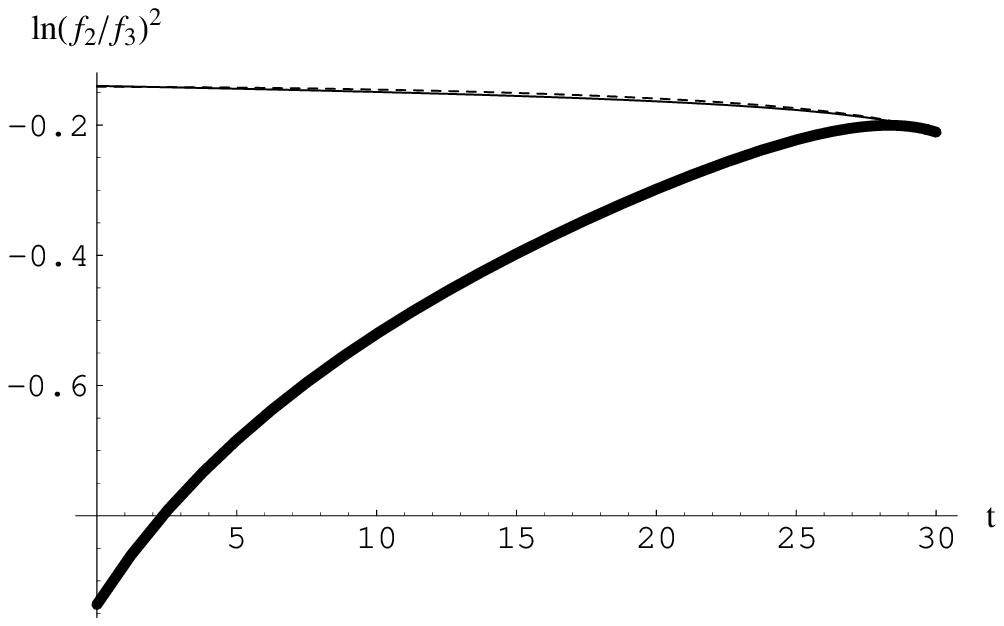} \\
    &(a)&\\
    \includegraphics[scale=0.7]{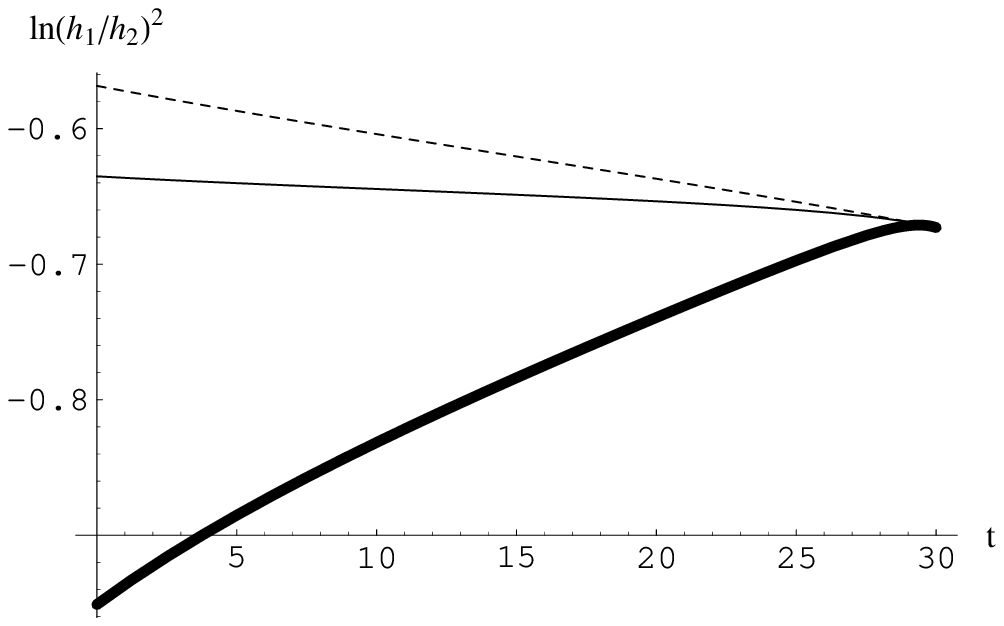} &&
    \includegraphics[scale=0.7]{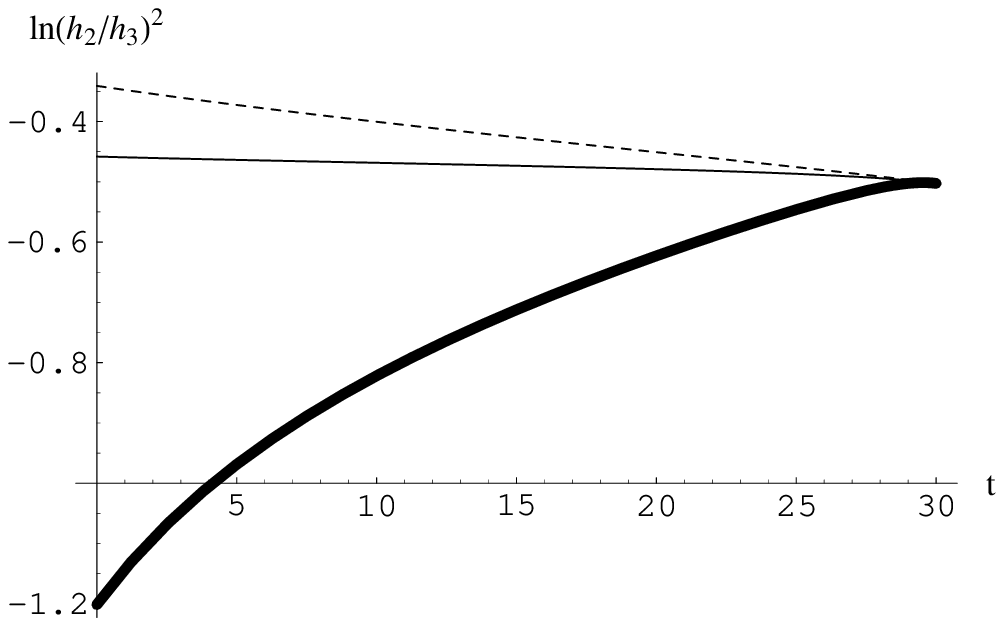} \\
    &(b)&
\end{tabular}
    \end{center}
\caption{The evolutions of (a) $f_i/f_j$ and (b) $h_i/h_j$ for SM(thin), MSSM(dashed) and the model(thick) with $(b,c)=(3,-7)$, with the boundary conditions [$(x_1,x_2)=(1/6+\epsilon,1/6-\epsilon)$,  $(y_1,y_2)=(-1/6+\epsilon,-1/6+\epsilon)$, $\epsilon=0.01$] and [$f_i=(1.6, 1.8, 2.0)$, $h_i=(0.5,0.7,0.9)$] at $t$=30.}\label{fig:mass ratios}
\end{figure}

Turning to the individual parameters, we present in Figs.(\ref{fig:xy}, \ref{fig:mass ratios}) the evolution of $(x_i,y_j)$ and the mass ratios. Using the same set of input initial values $(x_i\sim1/6,y_j\sim-1/6)$, we see that they behave similarly to products $\prod x_i$ and $\prod y_j$, albeit with slower rates of evolution. For the $(x_i,y_j)$ parameters, with $c=-7$, we find that the evolution occurs rapidly and almost saturates within $\Delta t\sim5-10$ from the initial $t$ value, with $x_1\rightarrow1$ and all the other $(x_i,y_j)\rightarrow0$. For $c=-3/2$(SM) and $c=1$(MSSM), we do not find such convergence. The evolution of the mass ratios behaves similarly, although even with $c=-7$, the hierarchy near $t=0$ does not approach the magnitude observed experimentally. Also, changing the $b$ values has only minimal influence on these general results. Despite the similarity of Eqs.(\ref{eq:DQ}) and (\ref{eq:Dq}) to Eqs.(\ref{eq:Dx}), (\ref{eq:Dy}) and (\ref{eq:DJ}), the numerical rates of evolution for the mass ratios differ considerably from those for the mixing parameters. A possible cause may be our choice of initial input values. However, so far our search for a region of input parameters which may result in a different behavior has not been successful.

In summary, we have used numerical analysis to explore the properties of the RGE for the $(x_i,y_j)$ parameters and the mass ratios. We assume that, at high energies, all of the $(x_i,y_j)$ values and the coupling constants $(f,h)$ are O(1), and that $(\Delta f,\Delta h)$ are small. We also take the coefficients $b$ and $c$ as free parameters. We find a physically interesting scenario if $c$ is negative and large. In this case, hierarchies in $(x,y)$ develop rapidly at high energies. It is expected that, at intermediate or low energies, the model will be superseded by other models (such as the SM or MSSM). However, the hierarchical values are close to a fixed point of the RGE, so we will not see much evolution in these energy regions. Thus, the observed hierarchy is primarily accomplished  at high energies, and will receive at most minor corrections at lower energies. At the same time, the hierarchies developed for the mass ratios are still not large enough to match those of the physical values. So the scenario of ``infrared hierarchy'' is only partially fulfilled. Further work is needed in order for one to arrive at a realistic model where the observed values of both $(x,y)$ and mass ratios can be reached within this framework.
\newpage
\newpage

\newpage
\section{Conclusion}\label{Sec:Conclusion}

In this paper we studied the RGE of the quark mass matrices, using the recently proposed flavor mixing parametrization which is manifestly rephasing invariant. The resulting set of equations is not as complicated as the one using different parametrizations. Also, the dependence on the variables entering the RGE is quite transparent and simple. We also presented, for some simple functions of mass ratios, evolution equations which bear close resemblance to those of the $(x,y)$ parameters. Since all of these quantities can be expressed as powers of $\lambda^2$, it is tempting to theorize that the hierarchies observed in mass ratios and in $(x,y)$ may both originate from renormalization effects. In fact, when we substitute their known physical values into RGE, it is seen that one is close to a fixed point and there is little evolution for these parameters. Given the uncertainties in the initial values and the theoretical models, extrapolation from low energies is unlikely to yield an accurate picture of renormalization effects. Rather, to assess the general nature of the RGE, it seems more appropriate to start from a point with fast evolution, so that most changes are accomplished in its neighborhood, with minor corrections afterwards. To this end we assume that, at high energies, the quark masses are large but nearly degenerate while $(x,y)$ values are not far from the ``symmetric point'', $x_i=-y_j=1/6$. It is found that, for a range of initial values, the parameters do evolve toward hierarchy. However, to reach the observed large hierarchy, one has to invoke the existence of new theories which give rise to large $c$ values contained in Eqs.(\ref{eq:DMu},\ref{eq:DMd}). On the other hand, even with these parameter choices, the mass ratios do not match the large hierarchy in the observed values. Thus, we may have established an outline of an ``infrared hierarchy'' scenario. The details, however, are still lacking. Hopefully, our analysis can provide the impetus for further researches along this direction.

\appendix
\appendixpage
\addappheadtotoc
\section{Properties of RGE Matrices}\label{Appx:AiBi}

In Eqs.(\ref{eq:Dx}, \ref{eq:Dy}), the RGE contain the matrices ($A_i$, $B_i$, $A^\prime_i$, $B^\prime_i$), which are listed explicitly in Table \ref{t:AiBi}. We will now discuss some of their properties in detail.

We note first that, since $\sum\Delta f_{ij}=\sum\Delta h_{ij}=0$, the RGE are invariant if a constant is added to each column of ($A_i$, $B_i$, $A^\prime_i$, $B^\prime_i$). For instance,
\begin{equation}
A_i\rightarrow A_i+\left(\begin{array}{ccl}
              \delta_1& \delta_2&\delta_3\\
              \delta_1&\delta_2&\delta_3\\
              \delta_1&\delta_2&\delta_3
            \end{array}\right)\label{eq:A-to-A+delta}
\end{equation}
leaves Eq.(\ref{eq:Dx}) invariant. The symmetric pattern of the matrices listed in Table \ref{t:AiBi} is thus not inherent and can be changed by transformations as in Eq.(\ref{eq:A-to-A+delta}). The constraints on the $(x,y)$ parameters imply the existence of consistency relations among the $(A,B)$ matrices. Indeed, it is found that $\sum(A_i-A^\prime_i)=\sum(B_i-B^\prime_i)=0$, which ensures $\mathscr{D}(\sum x_i-\sum y_i)=0$. However, these relations are valid only if in Eq.(\ref{eq:A-to-A+delta}) all $\delta_j=0$, $j=1,2,3$. To account for the arbitrary choice of $\delta_j$, we have the more general relations:
\begin{equation}
(1-P)\sum(A_i-A^\prime_i)=0,~~~(1-P)\sum(B_i-B^\prime_i)=0.\label{eq:identity-x+y}
\end{equation}
Here, $P$ is the cyclic permutation operator
\begin{equation}
P=\left(\begin{array}{ccl}
              0&0&1\\
              1&0&0\\
              0&1&0
            \end{array}\right),\label{eq:P-matrix}
\end{equation}
so that ($1-P$) annihilates the $\delta$ matrix in Eq.(\ref{eq:A-to-A+delta}), and $(\Delta f_{23},\Delta f_{31},\Delta f_{12})=(f^2_2,f^2_3,f_1^2)(1-P)$. One also establishes the relations
\begin{equation}
(1-P)\sum_{cyc}[(x_i+x_j)A_k-(y_i+y_j)A^\prime_k]=\sum_{i>j}(x_ix_j-y_iy_j)(1-P)w,\label{eq:identity-xyA}
\end{equation}
\begin{equation}
(1-P)\sum_{cyc}[(x_i+x_j)B_k-(y_i+y_j)B^\prime_k]=\sum_{i>j}(x_ix_j-y_iy_j)(1-P)w^T.\label{eq:identity-xyB}
\end{equation}
They validate the constraint $\mathscr{D}[\underset{i>j}{\sum}(x_ix_j-y_iy_j)]=0$. When we compute $\mathscr{D}(x_1x_2x_3)$ and $\mathscr{D}(y_1y_2y_3)$, we find
\begin{equation}
(1-P)\sum_{cyc}(x_ix_jA_k-y_iy_jA^\prime_k)=2J^2(1-P)w,\label{eq:identity-xyAJ}
\end{equation}
\begin{equation}
(1-P)\sum_{cyc}(x_ix_jB_k-y_iy_jB^\prime_k)=2J^2(1-P)w^T.\label{eq:identity-xyBJ}
\end{equation}
If we put these two equations together, we obtain the evolution equation for $J^2$, Eq.(\ref{eq:DJ}), given in Sec.\ref{Sec:RGE}. In the above equations, $\underset{cyc}{\sum}$ denotes summation over cyclic permutation of ($i,j,k$) so that, e.g., $\underset{cyc}{\sum}(x_i+x_j)A_k=(x_1+x_2)A_3+(x_2+x_3)A_1+(x_3+x_1)A_2$,  $\underset{cyc}{\sum}x_ix_jA_k=x_1x_2A_3+x_2x_3A_1+x_3x_1A_2$, etc.

The matrices ($A_i$, $B_i$, $A^\prime_i$, $B^\prime_i$) are also related to the $w$ matrix. Let us write
\begin{equation}
w=\bar{X}+\bar{Y},
\end{equation}
where
\begin{equation}
\bar{X}=\left(\begin{array}{ccc}
x_1&x_2&x_3\\
x_3&x_1&x_2\\
x_2&x_3&x_1
\end{array}\right),~~
\bar{Y}=\left(\begin{array}{ccc}
y_1&y_2&y_3\\
y_2&y_3&y_1\\
y_3&y_1&y_2
\end{array}\right).
\end{equation}
Then, e.g., the matrix $A_2$ can be written as
\begin{equation}
A_2=\left(\begin{array}{ccc}
x_1&y_2&x_3\\
x_3&x_1&y_1\\
y_3&x_3&x_1
\end{array}\right)
\left(\begin{array}{ccc}
x_2&0&0\\
0&x_2&0\\
0&0&x_2
\end{array}\right)
+\left(\begin{array}{ccc}
y_1&x_2&y_3\\
y_2&y_3&x_2\\
x_2&y_1&y_2
\end{array}\right)
\left(\begin{array}{ccc}
y_3&0&0\\
0&y_2&0\\
0&0&y_1
\end{array}\right).
\end{equation}
So, to obtain $A_2$, which is contained in $\mathscr{D}x_2$, we begin by exchanging the $x_2$ entries in $\bar{X}$ with corresponding ones in $\bar{Y}$, and then multiply these matrices by diagonal ones composed of the exchanged elements in $\bar{X}$ and $\bar{Y}$. Similar rules apply to the construction of the other matrices. Notice also that the matrices are related by exchanging $x_2\leftrightarrow x_3$:
\begin{equation}
\begin{array}{ccl}
A_1&\leftrightarrow&B_1;~~A_2\leftrightarrow B_3;~~A_3\leftrightarrow B_2;\\
A^\prime_1&\leftrightarrow&B^\prime_1;~~A^\prime_2\leftrightarrow B^\prime_2;~~A^\prime_3\leftrightarrow B^\prime_3.
\end{array}
\end{equation}
\section{Two-loop Renormailzation}\label{Appx:two-loop RGE}

The one-loop RGE, Eqs.(\ref{eq:DMu}, \ref{eq:DMd}), must be amended when higher order effects are considered. In particular, two-loop contributions can be included by writing
\begin{equation}
\mathscr{D}M_u=D^{(1)}_u+D^{(2)}_u,
\end{equation}
\begin{equation}
\mathscr{D}M_d=D^{(1)}_d+D^{(2)}_d.
\end{equation}
where $D^{(1)}_{u,d}$ are the one-loop contributions as given in Eqs.(\ref{eq:DMu}, \ref{eq:DMd}), and $D^{(2)}_{u,d}$ are given by
\begin{equation}
\begin{array}{ccl}
D^{(2)}_u&=&\{a^{(2)}_u+b^{(2)}_uM_u+c^{(2)}_uM_d
+d^{(2)}_uM^2_d+e^{(2)}_uM_uM_d+f^{(2)}_uM_dM_u+g^{(2)}_uM^2_u,M_u\},\\\\
D^{(2)}_d&=&\{a^{(2)}_d+b^{(2)}_dM_d+c^{(2)}_dM_u
+d^{(2)}_dM^2_u+e^{(2)}_dM_dM_u+f^{(2)}_dM_uM_d+g^{(2)}_dM^2_d,M_d\}.
\end{array}
\end{equation}
The coefficients $a$ through $g$ are given explicitly in Ref. \cite{Barger:1992ac} and will not be reproduced here.

Following the same procedure as in the one-loop calculations, we find the RGE for the $(x,y)$ parameters:
\begin{equation}
\begin{array}{ccl}
-\mathscr{D}^{(2)}x_i=\Delta \tilde{f}.A_i.H+\Delta \tilde{h}.B_i.F+\Delta f.A_i.\tilde{H}+\Delta h.B_i. \tilde{F},
\end{array}
\end{equation}

\begin{equation}
-\mathscr{D}^{(2)}y_i
=\Delta \tilde{f}.A^\prime_i.H+\Delta \tilde{h}.B^\prime_i.F+\Delta f.A^\prime_i.\tilde{H}+\Delta h.B^\prime_i.\tilde{F},
\end{equation}
where
\begin{equation}
\Delta \tilde{f}_{ij}=[c^{(2)}_d+d^{(2)}_d(f^2_i+f^2_j)]\Delta f_{ij},~~\Delta \tilde{h}_{ij}=[c^{(2)}_u+d^{(2)}_u(h^2_i+h^2_j)]\Delta h_{ij},
\end{equation}
\begin{equation} \tilde{F}_{ij}=\left[2e^{(2)}_u(f^2_if^2_j)+f^{(2)}_u(f^4_i+f^4_j)\right]/\Delta f_{ij},~~~ \tilde{H}_{ij}=\left[2e^{(2)}_d(h^2_ih^2_j)+f^{(2)}_d(h^4_i+h^4_j)\right]/\Delta h_{ij}.
\end{equation}

Note that $\sum\Delta \tilde{f}_{ij}=\sum\Delta \tilde{h}_{ij}=0$. Also, the structure of these equations is similar to that in Eqs.(\ref{eq:Dx}, \ref{eq:Dy}). In addition, the matrices $(A_i,A^\prime_i,B_i,B^\prime_i)$ enter these equations sandwiched between functions whose dependence on $\Delta f_{ij}$ and $\Delta h_{ij}$ is the same as in Eqs.(\ref{eq:Dx}, \ref{eq:Dy}). Thus, the effect of two-loop contributions may be approximated by a change in the value $c$ in Eqs.(\ref{eq:Dx}, \ref{eq:Dy}), especially in the neighborhood of degenerate masses.

\begin{acknowledgments}
S.H.C. is supported by the National Science Council of Taiwan under Grant No. NSC 97-2112-M-182-001. T.K.K. and C.X. are supported by the Grant at Purdue [DE-FG02-91ER40681(task B)]. C.X. is also supported, in part, by the U.S. Department of Energy under Grant No. DE-FG02-97ER41027.
\end{acknowledgments}

\end{document}